\documentclass[12pt]{article}
\usepackage{graphicx}
\begin{document}

\title{Incomplete information and fractal phase space}

\author{Qiuping A. Wang \\ Institut Sup\'erieur des Mat\'eriaux du Mans, \\
44, Avenue F.A. Bartholdi, 72000 Le Mans, France}

\date{}

\maketitle

\begin{abstract}
The incomplete statistics for complex systems is characterized by a so called incompleteness parameter $\omega$
which equals unity when information is completely accessible to our treatment. This paper is devoted to the
discussion of the incompleteness of accessible information and of the physical signification of $\omega$ on the
basis of fractal phase space. $\omega$ is shown to be proportional to the fractal dimension of the phase space
and can be linked to the phase volume expansion and information growth during the scale refining process.
\end{abstract}

{\small PACS number : 02.50.Cw,02.70.Rr,89.70.+c,89.75.Da}

\vspace{1cm}

{\bf Introduction}

A fundamental hypothesis in Boltzmann statistical physics, explicit or not, is that the probability
distribution is complete so that $\sum_{i=1}^{w}p_i=1$, where $w$ is the number of all the possible states
labelled by $i$ and $p_i$ is the probability that the system is at the state $i$. This is the {\it complete
probability normalization}\cite{Reny66}. In other words, one can treat a complete or asymptotically complete
set of states of the system under consideration. This hypothesis of complete statistics implies that the
following two conditions must be satisfied. 1) All possible states of the system of interest are well known,
which requires a complete knowledge of the dynamics of the system. 2) The known states are accountable and
calculable, or in other words, the information is accessible {\it to us}. Are these two conditions always
satisfied?

As the study of complexity advanced, it is revealed that chaotic and fractal behaviors are ubiquitous in
nature\cite{Hilb94,Ruelle91,Thuan,Complex,Prants}, and that a fractal phase space in general cannot be
completely and exactly exploited. Any calculation based only on the accessible states or the differentiable
(integrable) points in this phase space is necessarily incomplete due to the rejected or unknown (singular or
unaccessible) points. So the complete descriptions may be satisfactory if and only if these unaccessible points
are negligible. If it is not the case, these points should be taken into account or the theoretical
descriptions would be aberrant. The question is : how to include the states of a physical system in a theory
when they are not accessible to us?

Recently, we have developed a method in this direction\cite{Wang01,Wang01a,Wang03,Wang02d,Wang02e,Wang02a} : a
statistical mechanics based upon the notion of {\it incomplete information} which has been shown capable of
giving $coherently$ the Tsallis $q$-exponential distributions within the nonextensive statistical mechanics
(NSM)\cite{Wang01} and, moreover, useful quantum distributions for some systems of correlated
electrons\cite{Wang01a,Wang02d}. In the present paper, I am presenting some arguments to support the hypothesis
of information incompleteness. The physical significations of the incompleteness parameter will be discussed on
the basis of fractal phase space.

\vspace{0.5cm}

{\bf Incomplete information hypothesis}

In the conventional probabilistic science, it is supposed that our ignorance, or information, is $completely$
accessible and may be calculated by, e.g., the Hartley formula $I_i=\ln(1/p_i)$ and Shannon information measure
$I=\sum_i^wp_i\ln(1/p_i)$\cite{Shannon,Hartley}, as well as by other generalized information
measures\cite{Reny66,Tsal88,Cura91,Tsal99,Penni}, under the {\it harsh} condition that all possible states are
accessible to us so that all probabilities sum to one.

However, the hypothesis of {\it incomplete information} admits simply that a part of our ignorance about
complex system may not be accessible to our treatment. It cannot be calculated from probability distributions.
In other words, the information calculated by the above mentioned method on the basis of the complete
probability normalization may be incomplete because here we only have $\sum_{i=1}^{v}p_i=\Omega$\cite{Reny66}
where $v$ is the number of the accessible or accountable states and may be greater or smaller than $w$, the
total number of states. $\Omega$ represents the incompleteness of the treatment and necessarily linked to the
nature of the system. Logically, the origine of this incompleteness may be attributed either to the partial
knowledge of the dynamics or to the unaccessible (incalculable) states of the system.

Incomplete information hypothesis has been motivated originally by some fundamental problems encountered in
NSM\cite{Wang01,Wang01a,Wang02d,Wang02e,Wang02a,Tsal88,Cura91,Tsal99,Penni}. It replaces the {\it complete
probability normalization} by\cite{Wang01}
\begin{equation}                                            \label{1}
\sum_{i=1}^{v}p_i^\omega=1,
\end{equation}
where $\omega$ is referred to as {\it incompleteness parameter} which equals unity if the probability
distribution is complete\cite{Reny66}. Then on the basis of the Hartley additive information measure
$I_i=\ln(1/p_i)$ and the nonadditive generalization of Hartley formula
$(I_i)_\omega=\ln_\omega(1/p_i)=\frac{(1/p_i)^{\omega-1}-1}{\omega-1}$, two incomplete entropies $S_{\omega
1}=-k\sum_ip_i^\omega\ln p_i$ and $S_{\omega 2}=k\frac{1-\sum_ip_i}{1-\omega}$ $(\omega \in
R)$\cite{Wang01,Wang01a} are deduced. These formalisms not only give in a coherent way the nonextensive
incomplete distributions for canonical ensemble $p_i\propto[1-(1-\omega)x]^{\omega/(1-\omega)}$ which has been
proved very useful for many systems having non-gaussian distributions, they also lead to quantum distributions
of which the first results of the applications to correlated electron systems seem quite interesting and
promising\cite{Wang01a,Wang02d,Wang02a}.

With the {\it incomplete normalization} Eq.(\ref{1}), the parameter $\omega$ can be uniquely related to the
incompleteness $\Omega$ by
\begin{equation}                                            \label{2}
\sum_{i=1}^{v-1}p_i^\omega+(\Omega-\sum_{i=1}^{v-1}p_i)^\omega=1,
\end{equation}
for any $0<p_i<1$, $v$ and $\Omega$. Figure 1 shows the relation between $\omega$ and $\Omega$ for a given
incomplete distribution $\{p_{i=1...5}\}=\{0.1, 0.15, 0.2, 0.25, \Omega-0.7\}$. We see that, in general,
$\Omega>1$ and $\Omega<1$ lead to $\omega>1$ and $\omega<1$, respectively. If $\Omega$ is such that a certain
$p_k\rightarrow 1$ among $\{p_{i=1...5}\}$, then $\omega\rightarrow\infty$. If $\Omega\rightarrow 0$,
$\omega\rightarrow 0$.

\begin{figure}[h] \label{f1}
\includegraphics[width=4in,height=3in]{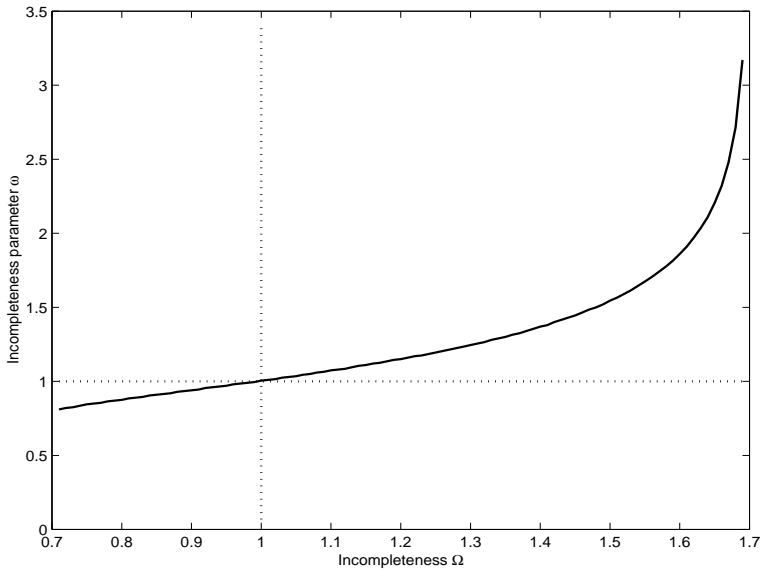}
\caption{$\omega-\Omega$ dependence for a given distribution $\{p_{i=1...5}\}=\{0.1, 0.15, 0.2, 0.25,
\Omega-0.7\}$. Note that $\omega\rightarrow \infty$ when $\Omega\rightarrow 1.7$ in this case.}
\end{figure}

\vspace{0.5cm}

{\bf Incomplete normalization in its general form}

In general, we should write
\begin{equation}                                            \label{3}
\sum_{i=1}^vF_\omega(p_i)=1
\end{equation}
in order to normalize incomplete distributions. The function $F_\omega$ should depend on the nature of the
system and become identity function whenever information is supposed complete ($\Omega=1$). The arithmetic
average of a quantity $\mathbf{x}$ should now be given by
\begin{equation}                                            \label{4}
\bar{x}=\sum_{i=1}^vF_\omega(p_i)x_i.
\end{equation}
where $x_i$ is the value of $\mathbf{x}$ at the state $i$.

$F_\omega$ may be determined if the information measure and the distribution law are given. For example, with
Hartley information measure and exponential distribution, $F_\omega$ can be showed to be identity
function\cite{Wang01a}. In general, by entropy maximization through the functional
\begin{equation}                                            \label{5}
\delta [\sum_iF_\omega(p_i)I(p_i)+\alpha\sum_iF_\omega(p_i)+\beta\sum_iF_\omega(p_i)x_i]=0
\end{equation}
we get :
\begin{equation}                                            \label{6}
\frac{\partial \ln F_\omega(p_i)}{\partial p_i}=\frac{\partial I/\partial p_i}{I+\alpha+f_\omega^{-1}(p_i)}
\end{equation}
or $F_\omega(p_i)=C\exp[\int\frac{\partial I/\partial p_i}{I+\beta f_\omega^{-1}(p_i)}dp_i]$
where $\alpha$ and $\beta$ are the multipliers of Lagrange in Eq.(\ref{5}), $I(p_i)$ is the
information measure, $p_i=f_\omega(x_i)$ the distribution function depending on the parameter
$\omega$, and $C$ the normalization constant of $F_\omega$.

Within this frame, we would be able to group many entropies\cite{Esteban} into the family of
incomplete entropies if we suppose that the Hartley formula $I_i=\ln(1/p_i)$ holds. For
example, the additive entropy proposed by Gorban and coworkers\cite{Gorb} to generalize
Gibbs-Shannon entropy and to obtain the long and short tail distributions may be considered as
the expectation of Hartley information with $F_\alpha(p_i)=p_i-\alpha(p_i+1)$ for an
incomplete description discussed by the authors. In this case, the incompleteness of the
description may be given by $\sum_{i=1}^{v}p_i=\frac{1-\alpha v}{1-\alpha}$. When $\alpha=0$,
the description becomes complete and Gorban entropy recovers Gibbs-Shannon one. The second
example is the entropy due to Burg for solving geophysical problems\cite{Burg} : $S \propto
\sum_i\ln p_i$. To have this, we may put $F_\omega(p_i)=1/v$ which is a
pseudo-equiprobability. Other examples are the entropies of Belis-Guiasu\cite{Belis} with
$F_\omega(p_i)=w_ip_i$ for the study of the information in cybernetic systems and the
entropies of Aczel-Daroczy\cite{Daroczy} with $F_\omega(p_i)=p_i^\gamma$. So the incomplete
normalization seemingly provides a plausible approach to link certain special entropies to
elementary information.

\vspace{0.5cm}

{\bf About incomplete information in fractal phase space}

Through a detailed analysis of the entropy and information on fractal supports, Naschie\cite{Naschie1} shows a
connection between the information and the topological dimension of some fractal sets. It is also shown that
the dynamics of this kind of phase space is quasi-ergodic for $n\geq 4$ where $n$ is the dimension of the set
or the number of particles in the fractal phase space of topological dimension $d_f$\cite{Naschie1,Naschie2}.
The results of the analysis tell us that the Boltzmann entropy and Gibbs-Shannon entropy for Cantor set
increase with increasing capacity dimension given by $d_c^{(n)}=(1/d_f)^{n-1}$ which in turn increases with
increasing dimension $n$ of the fractal set. On the other hand, $d_c^{(n)}$ should increase with decreasing
topological dimension $d_f$. This behavior is similar to that of $S_{\omega 1}$ and $S_{\omega 2}$ with respect
to the parameter $\omega$\cite{Wang01,Wang03}, which means that $\omega$ may have something to do with the
topological dimension in fractal phase space.

In what follows, the information on fractal support will be studied from a different angle with the help of the
incomplete normalization Eq.(\ref{1}). First of all, we will show that this kind of power normalization is
inevitable for the calculation of fractal information.

\begin{figure}[p] \label{f2}
\includegraphics[width=4in,height=3in]{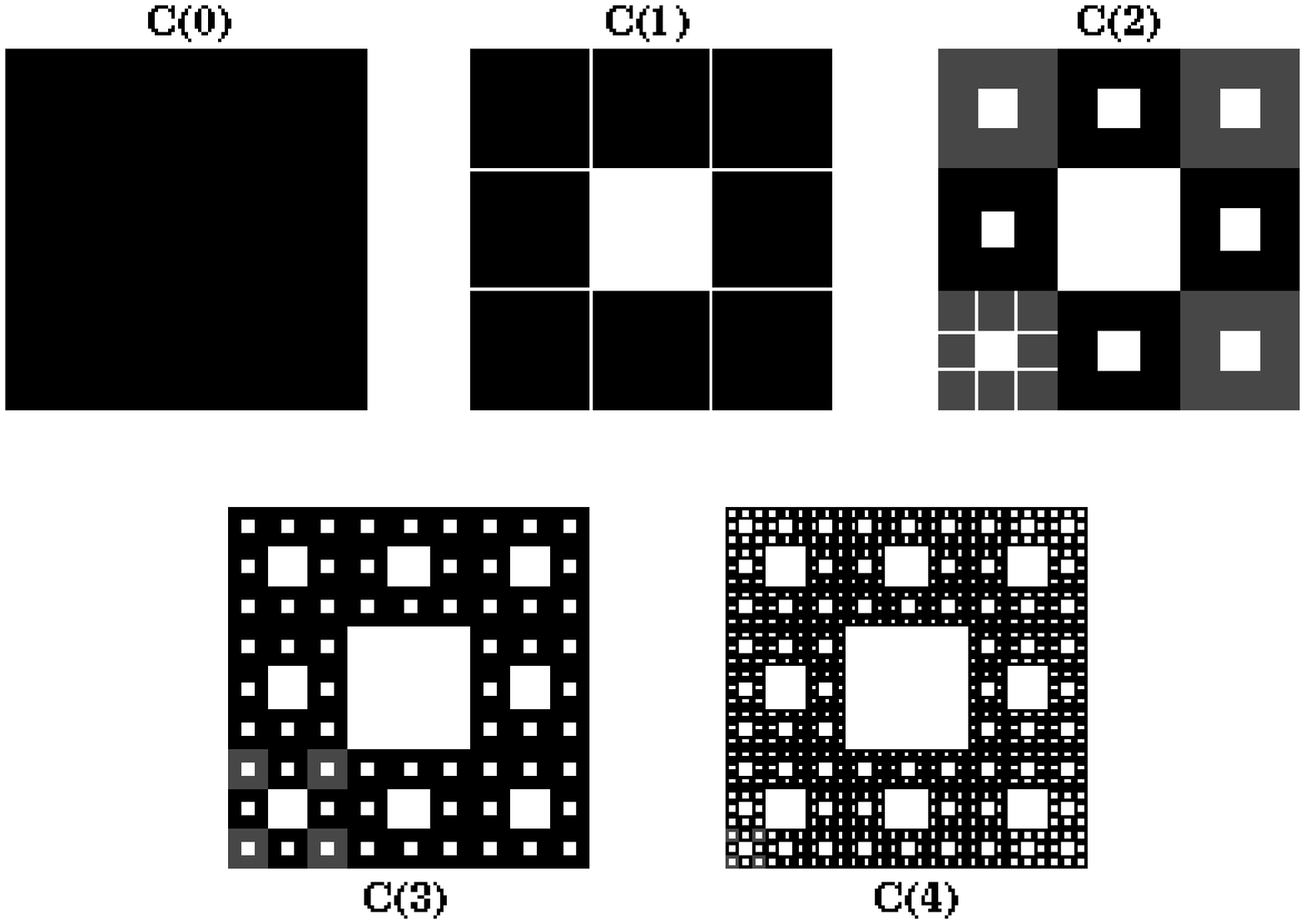}
\caption{A simple model of fractal phase space in Sierpinski carpet which is a self-similar
fractal structure generated by dividing, at the first iteration, a full square of side $l_0$
into $3^2$ smaller squares of equal sides of $l_0/3$ and removing the central small square.
This procedure will be repeated on every small square at the next iteration and so on. At
$k^{th}$ iteration, the side of the squares (black or white) is $l_k=l_0/3^k$ and their number
is $W_k=8^k$, $l_0$ being the length of the side at $0^{th}$ iteration. The total surface at
$k^{th}$ iteration is $S_k=W_k s_k$ or  $W_k s_k/S_k=1$. The classical probability definition
by relative frequency of visits of each point by the system should be modified because the
total number of visits (proportional to black surface $S_k$ of the carpet) is no more a finite
quantity. (Construction of Sierpinski carpet. First iteration c(1) : removing the central
square formed by the straight lines cutting each side into three segments of equal size.
Repeat this operation on the 8 remaining squares of equal size and so on.)}
\end{figure}

\vspace{0.5cm}

{\bf An incomplete normalization }

For the sake of simplicity, let us consider a phase space in which the trajectory of a complex system forms a
self-similar fractal structure, say, Sierpinski carpet (Figure 1). This means that the state point of the system
can be found only on the black rectangular segments whose number is $W_k=8^k$ at $k^{th}$ iteration. Hence the
total surface at this stage is given by $S_k=W_k s_k$ where $s_k=l_0/3^k$ is the surface of the segments at
$k^{th}$ iteration and $l_0$ the length of side of the square space at $0^{th}$ iteration. If the segments do
not have same surface, we should write $S_k=\sum_{i=1}^{W_k}s_k(i)$.

Now let us suppose that the density of state is identical everywhere on the segments and that the dynamics in
this phase space is quasi-ergodic : {\it every point on the segments is equally visited}, so that the
probability for the system to be in the $i^{th}$ segment may be defined as usual by $p_i=s_k(i)/S_k$. This
probability is obviously normalized. The problem is that, as discussed in \cite{Hilb94}, $S_k$ is an indefinite
quantity as $k\rightarrow \infty$ or as the time $t\rightarrow \infty$. So strictly speaking, it can not be
used to define exact probability definition or all the probabilities would be null or lose its additivity and
normalization according to the countable additivity axiom of probability or total probability axiom which
asserts that probabilities can be additive on a set of finite cardinals (number of elements). In addition,
$S_k$ is not differentiable and contains inaccessible points. Thus the probability defined above makes no
sense.

Alternatively, the probability may be reasonably defined on a integrable and differentiable support, say, the
Euclidean space containing the fractal structure. To see how to do this, we write
$S_k=l_0^2(\frac{1}{3^k})^{d-d_f}$ for identical segments or, for segments of variable size,
\begin{equation}                                \label{8}
\sum_{i=1}^{W_k}[\frac{s_k(i)}{S_0}]^{d_f/d}=1
\end{equation}
where $S_0=l_0^d$ (here $d=2$ for Sierpinski carpet) is a characteristic volume of the fractal structure
embedded in a $d$-dimension Euclidean space, $d_f=\frac{\ln n}{\ln m}$ is the fractal dimension. $n=8$ is the
number of segments replacing a segment of the precedent iteration and $m=3$ the scale factor of the iterations.

The microcanonical probability distribution at the $k^{th}$ iteration can be defined as
$$p_i=\frac{s_k(i)}{S_0}$$ so that $$\sum_{i=1}^{W_k}p_i^{d_f/d}=1.$$ This is just Eq.(\ref{1}) with
$\omega=d_f/d$. The complete probability normalization $\sum_{i=1}^{W_k}p_i=1$ can be recovered when $d_f=d$.

It should be noticed that, in Eq.(\ref{8}), the sum over all the $W_k$ segments at the $k^{th}$ iteration does
not mean the sum over all possible states of the system under consideration. This is because that the segment
surface $s_k(i)$ does not represent the real number of state points on the segment which, as expected for any
self-similar structure, evolves with $k$ just as $S_k$. So at any given order $k$, the complete summation over
all possible segments is not a complete summation over all possible states. But in any case, whatever is $k$,
Eq.(\ref{8}) and $\sum_{i=1}^{W_k}p_i^\omega=1$ always hold for $\omega=d_f/d$.

In this simple case with self-similar fractal structure, the incompleteness of the normalization Eq.(\ref{1})
is measured by the parameter $\omega=d_f/d$. $d_f>d$ means that the system has more states than $W_k$, the
number of accessible states at given $k$. If $d_f<d$, the number of states is less than $W_k$. When $d_f=d$,
the summation is complete at any order $k$, corresponding to complete information calculation.

\vspace{0.5cm}

{\bf $\omega$ and phase space expansion}

Now we will discuss in a detailed way the incompleteness parameter $\omega$ and its physical meanings. Let us
discuss this still in the case of self-similar fractal phase space.

As discussed in the case of chaotic phase space, $\omega=\ln n/d\ln m$ gives a measure of the incompleteness of
the state counting in the $d$-dimension phase space. $\omega=1$ means $d_f=d$ or $n=m^d$. In other word, at the
$k^{th}$ iteration, a segment of volume $s_k$ is completely covered (replaced) by $n$ segments of volume
$s_{k+1}=s_k/m^d$. So the summation over all segments is equivalent to the sum over all possible states, making
it possible to calculate complete information.

When $\omega>1$ (or $\omega<1$), $n>m^d$ (or $n<m^d$) and $s_k$ is replaced by $n$ segments whose total volume
is more (or less) than $s_k$. So there is expansion (or negative expansion) of state volume when we refine the
phase space scale. An estimation of this expansion at each scale refinement can be given by the ratio
$r=\frac{ns_{k+1}-s_k}{s_k} =\frac{n}{m^d}-1 =(\frac{1}{m^d})^{1-\omega}-1 =
(\omega-1)\frac{(m^d)^{\omega-1}-1}{\omega-1}$. $r$ describes {\it how much accessible states increase} at each
step of the iteration or of the refinement of phase space. The physical content of $\omega$ is clear if we note
that $\omega>1$ and $\omega<1$ correspond to an expansion ($r>0$) and a negative expansion ($r<0$),
respectively, of the the accessible state volume at each step of the iteration.

When $\omega=0$, we have $d_f=0$ and $n=1$, leading to $r=\frac{1}{m^d}-1$. The iterate condition $n\geq 1$
means $\omega\geq 0$, as proposed in references \cite{Wang01}. $\omega<0$ is impossible since it means $d_f<0$
or $n<1$ which obviously makes no sense. We can also write : $\omega-1=\ln(r+1)/\ln(m^d)
=\ln(ns_{k+1}/s_k)/\ln(m^d)$, which implies that it is the difference $\omega-1$ which directly measures the
accessible state space expansion through the scale refinement.

\vspace{0.5cm}

{\bf $\omega$ and information growth}

The expansion of the accessible state volume of a system in its phase space during the scale refinement should
be interpreted as follows : the extra state points $\Delta=ns_{k+1}-s_k$ acquired at $(k+1)^{th}$ order iterate
are just the number of unaccessible states at $k^{th}$ order with respect to $(k+1)^{th}$ order. $\Delta>0$ (or
$\Delta<0$) means that we have counted less (or more) states at $k^{th}$ order than we should have done.
$\Delta$ contains the {\it accessible information gain} (AIG) through the $(k+1)^{th}$ iterate.

To illustrate the relation between this ``hidden information" and the parameter $\omega$, let us suppose that
{\it the distribution is scale-invariant}\cite{Invariant}. At the iterate of order $k$, the average information
contained on $s_k$ is given by $I_k=\int_{s_k} p^\omega I(1/p)ds$. At $k+1$ order, $I_{k+1}=\int_{ns_{k+1}}
p^\omega I(1/p)ds$. Hence AIG is just $\Delta I=I_{k+1}-I_k=\int_{(ns_{k+1}-s_k)} p^\omega I(1/p)ds
=\sigma_I\Delta$, where $\sigma_I=p^\omega I(1/p)$ is the information density or the average information
carried by each state. The relative AIG is given by $\Delta
I/I_k=r=(1-\omega)\frac{(1/m^d)^{1-\omega}-1}{1-\omega}$ which is independent of scale but dependent on scale
changes. For given scaling factor $m$, the magnitude of $\Delta I$ or $r$ increases with increasing difference
$|1-\omega|$. The sign of $r$ (or AIG) was discussed earlier. For given $\omega$, $|\Delta I|$ increases with
decreasing scaling. For $\omega=1$ or $m=1$, there is no information gain, corresponding to the case of
complete information. According to the above discussions, the incompleteness parameter $\omega$ may be
considered as a measure of the degree of chaos or fractal.

\vspace{0.5cm}

{\bf Conclusion}

Summing up, we have argued that the information of a complex system we deal with may be incomplete because we
only partially know the dynamics of the system or a part of the states of the systems is not accessible to our
calculation. The parameterized normalization $\sum_ip_i^\omega=1$ is justified on fractal support, where the
incompleteness parameter $\omega$ is shown to be proportional to the fractal dimension of the phase space and
can be directly linked to the phase volume expansion and information growth during the scale refining process.

{\bf Acknowledgments}

I acknowledge with great pleasure the very useful discussions with Professor M.S. El Naschie.


\begin{thebibliography}{99}
\bibitem {Quasi}
By quasi-deterministic theory, I mean conventional probabilistic sciences which, although incapable of giving
deterministic description as does Newtonian mechanics, give nevertheless a complete description based on a
perfect knowledge of all possible states of the system of interest. We are not sure of the state of the system
at any given moment, but we know that the system must be at one of the states in our register given by the
equation of motion.
\end{thebibliography}
\end{document}